\documentclass[12pt,thmsa]{article}
\usepackage{sw20aip}

\input{tcilatex}
\begin{document}

\author{Emilio Santos}
\title{A local hidden variables model for the measured EPR-type flavour
entanglement in $Y\left( 4S\right) \rightarrow B^{0}\overline{B^{0}}$ decays }
\date{March, 15, 2007 }
\maketitle

\begin{abstract}
A local hidden variables model is exhibited which gives predictions in
agreement with the quantum ones for the recent experiment by Go et al.,
quant-ph/0702267 (2007)
\end{abstract}

The aim of this note is to show that the results of the recent experiment
measuring EPR-type flavour entanglement in $Y\left( 4S\right) \rightarrow
B^{0}\overline{B^{0}}$ decays\cite{Go} are compatible with local realism. It
is known that a Bell inequality test cannot be performed\cite{Bramon} but
this does not prove that the experiment is compatible with local realism. I
shall prove the compatibility by exhibiting a local hidden variables (LHV)
model which reproduces the quantum prediction (and agrees with the obtained
results within experimental errors). The time-dependent rate for decay into
two flavour-specific states are\cite{Go} 
\begin{equation}
R_{i}\left( \triangle t\right) =\frac{1}{4\tau }\exp (-\triangle t/\tau
)\left( 1+\left( -1\right) ^{i}\cos \left( \triangle m\triangle t\right)
\right) ,\;\triangle t=\left| t_{2}-t_{1}\right| ,  \label{1}
\end{equation}
where $\triangle m$ is the mass difference between the two $B^{0}-\overline{%
\text{ }B^{0}}$ mass eigenstates, $i=1$ corresponds to the decays $%
B^{0}B^{0} $ or $\overline{B^{0}}\overline{B^{0}}$ and $i=2$ to the decays $%
B^{0}\overline{B^{0}}$ or $\overline{B^{0}}B^{0}$. Actually $R_{i}$ are the
probability densities of decay at time $t_{2}$ of the second particle (say
the one going to the right) conditional to the decay of the first (say going
to the left) at time $t_{1}$, both $t_{1}$ and $t_{2}$ being proper times of
the corresponding particles. For our purposes it is more convenient to
consider the joint probability densities, $r_{kl}\left( t_{1},t_{2}\right) $
for decay of the first particle at time $t_{1}$ and second at time $t_{2}$,
where $k=1\;(l=1)$ means that the first (second) particle decays as $B^{0}$
and $k=2\;(l=2)$ means that the first (second) particle decays as $\overline{%
B^{0}}.$

According to Bell\'{}s definition of LHV model\cite{Bell}, appropriate for
our case, we should attach hidden variables $\lambda _{1}$ and $\lambda _{2}$
to the first and second particles, respectively, and define probability
densities $\rho ,P_{k},Q_{l}$ such that 
\begin{equation}
r_{kl}\left( t_{1},t_{2}\right) =\int \rho \left( \lambda _{1},\lambda
_{2}\right) P_{k}\left( \lambda _{1},t_{1}\right) Q_{l}\left( \lambda
_{2},t_{2}\right) d\lambda _{1}d\lambda _{2}.  \label{3}
\end{equation}
The function $\rho ,$ giving the initial distribution of the hidden
variables in an ensemble of $Y\left( 4S\right) $ decays, should be positive
and normalized, that is 
\begin{equation}
\rho \left( \lambda _{1},\lambda _{2}\right) \geq 0,\int \rho \left( \lambda
_{1},\lambda _{2}\right) d\lambda _{1}d\lambda _{2}=1.  \label{3a}
\end{equation}
The functions $P_{k}\left( \lambda _{1},t_{1}\right) $ represent the
probability density that a particle with label $\lambda _{1}$ decays at time 
$t_{1}$ as a $B^{0}\left( \overline{B^{0}}\right) $ if $k=1(2)$ and similar
for $Q_{l}.$ Thus these functions should be positive and, as all $B^{0}$ or $%
\overline{B^{0}}$ particles decay sooner or later, they should be normalized
for any $\left\{ \lambda _{1},\lambda _{2}\right\} $, that is 
\begin{equation}
P_{k},Q_{l}\geq 0,\int_{0}^{\infty }dt_{1}\sum_{k=1}^{2}P_{k}\left( \lambda
_{1},t_{1}\right) =1,\int_{0}^{\infty }dt_{2}\sum_{l=1}^{2}Q_{l}\left(
\lambda _{2},t_{2}\right) =1.  \label{3b}
\end{equation}
Any choice of functions $\{\rho ,P_{k},Q_{l}\}$ fulfilling eqs.$\left( \ref
{3}\right) $ to $\left( \ref{3b}\right) $ provides a LHV model predicting
the joint probabiliy densities of decay $r_{kl}\left( t_{1},t_{2}\right) .$

I propose the following. For the initial distribution of hidden variables 
\begin{equation}
\rho \left( \lambda _{1},\lambda _{2}\right) =\frac{1}{4\tau N\left( \lambda
_{2}\right) }\delta \left( \lambda _{1}-\lambda _{2}\right) ,\;\lambda
_{1},\lambda _{2}\in [0,2\pi ],  \label{4}
\end{equation}
where $\delta \left( {}\right) $ is Dirac\'{}s delta, and the functions $%
N\left( \lambda _{2}\right) $ will be defined below, after eqs.$\left( \ref
{4c}\right) ,$ where the normalization of $\rho \left( \lambda _{1},\lambda
_{2}\right) $ will be proved. For the probabilities of decay 
\begin{equation}
P_{k}\left( \lambda _{1},t_{1}\right) =\frac{1}{\tau }\exp \left( -\frac{%
t_{1}}{\tau }\right) \Theta \left( t_{1}\right) \sum_{n=0}^{\infty }\Theta
\left( \frac{\pi }{2}-\left| \lambda _{1}+(2n-k)\pi -\triangle mt_{1}\right|
\right) ,  \label{4a}
\end{equation}
\begin{equation}
Q_{l}\left( \lambda _{2},t_{2}\right) =N\left( \lambda _{2}\right) \exp
\left( -\frac{t_{2}}{\tau }\right) \Theta \left( t_{2}\right) \left[ \cos
\left( \lambda _{2}-(l+1)\pi -\triangle mt_{2}\right) \right] _{+},
\label{4d}
\end{equation}
where $\Theta \left( t\right) =1\;(0)$ if $t>0\;(t<0)$ and $\left[ x\right]
_{+}$ means putting $0$ if $x<0$. Thus all four functions are decaying
exponentials modulated by periodic funcions which oscillate with period $%
2\pi /\triangle m.$ Physically this means that each particle ``lives'' as a $%
B^{0}$ during a time interval of duration $\pi /\triangle m$, then becomes a 
$\overline{\text{ }B^{0}}$ during another time interval $\pi /\triangle m$,
and so on, until it decays. The particles are always anticorrelated in the
sense that, at equal proper times, one of them is $B^{0}$ and the other one
is $\overline{\text{ }B^{0}}.$

From eqs.$\left( \ref{4a}\right) $ and $\left( \ref{4d}\right) $ it is easy
to see that the total probability densities $\left( \text{i. e.
independently of flavour}\right) $ for the decay of particles 1 and 2, are
respectively 
\begin{eqnarray}
\sum_{k=1}^{2}P_{k}\left( \lambda _{1},t_{1}\right) &=&\frac{1}{\tau }\exp
\left( -\frac{t_{1}}{\tau }\right) ,  \nonumber \\
\sum_{l=1}^{2}Q_{l}\left( \lambda _{2},t_{2}\right) &=&N\left( \lambda
_{2}\right) \exp \left( -\frac{t_{2}}{\tau }\right) \left| \cos \left(
\lambda _{2}-\triangle mt_{2}\right) \right| ,  \label{4c}
\end{eqnarray}
where $t_{1},t_{2}\geq 0.$ We see that the decay of the first particle is
given by a standard exponential, but the decay law of the second particle is
more involved. The functions $N\left( \lambda _{2}\right) $ are chosen so
that the normalization eq.$\left( \ref{3b}\right) $ holds true. It is not
necessary to calculate explicitly the functions $N\left( \lambda _{2}\right) 
$, which are rather involved, but I derive an important property, namely 
\begin{eqnarray}
\int_{0}^{2\pi }\frac{1}{N\left( \lambda \right) }d\lambda &=&\int_{0}^{2\pi
}d\lambda \int_{0}^{\infty }\exp \left( -\frac{t}{\tau }\right) \left| \cos
\left( \lambda -\triangle mt\right) \right| dt  \nonumber \\
&=&\int_{0}^{\infty }\exp \left( -\frac{t}{\tau }\right) dt\int_{0}^{2\pi
}\left| \cos \left( \lambda -\triangle mt\right) \right| d\lambda =4\tau .
\label{4f}
\end{eqnarray}
This relation proves that the distribution $\rho \left( \lambda _{1},\lambda
_{2}\right) ,$ eq.$\left( \ref{4}\right) ,$ is indeed normalized.

In order to get $r_{kl}\left( t_{1},t_{2}\right) $ we should insert eqs.$%
\left( \ref{4a}\right) $ and $\left( \ref{4d}\right) $ in eq.$\left( \ref{3}%
\right) $ and perform integrals which are straightforward. Introducing the
new variable 
\begin{equation}
x=\lambda _{1}+2n\pi -k\pi -\triangle mt_{1},  \label{5a}
\end{equation}
and performing the integral in $\lambda _{2},$ using Dirac\'{}s delta, we
get 
\begin{eqnarray}
r_{kl}\left( t_{1},t_{2}\right) &=&\frac{1}{4\tau ^{2}}\exp \left( -\frac{%
t_{1}+t_{2}}{\tau }\right) I_{kl},  \nonumber \\
I_{kl} &=&\int_{-\pi /2}^{\pi /2}dx\left[ \cos \left( x+\left( k-l-1\right)
\pi +s\right) \right] _{+},\,s\equiv \triangle m(t_{1}-t_{2}),  \label{6}
\end{eqnarray}
where we have taken into account that only one term of the sum in $n$ may
contribute, depending on the values of $t_{1}$ and $t_{2}$, and we have
removed the irrelevant term $2n\pi $ in the argument of the cosinus
function. It is easy to see that the functions $I_{kl}$ are periodic in the
variable $s$ with period $2\pi .$ Thus it is enough to consider the interval 
$s\in \left[ 0,2\pi \right] .$ Thus in the particular cases $k=l=1$ or $%
k=l=2 $ the integral $\left( \ref{6}\right) $ becomes, for $s\in \left[
0,\pi \right] $ 
\begin{equation}
I_{11}\left( t_{1},t_{2}\right) =I_{22}\left( t_{1},t_{2}\right) =\int_{\pi
/2-s}^{3\pi /2-s}\left[ \cos x\right] _{+}dx=\int_{\pi /2-s}^{\pi /2}\cos
xdx=1-\cos s,  \label{7}
\end{equation}
and for $s\in \left[ \pi ,2\pi \right] $ 
\begin{equation}
I_{11}\left( t_{1},t_{2}\right) =I_{22}\left( t_{1},t_{2}\right) =\int_{\pi
/2-s}^{3\pi /2-s}\left[ \cos x\right] _{+}dx=\int_{-\pi /2}^{3\pi /2-s}\cos
xdx=1-\cos s.  \label{7a}
\end{equation}
Similarly we get, for any $s=\triangle m\,(t_{1}-t_{2}),$%
\begin{equation}
I_{12}\left( t_{1},t_{2}\right) =I_{21}\left( t_{1},t_{2}\right) =1+\cos s,
\label{7b}
\end{equation}
Finally we obtain 
\begin{equation}
r_{kl}\left( t_{1},t_{2}\right) =\frac{1}{4\tau ^{2}}\exp \left( -\frac{%
t_{1}+t_{2}}{\tau }\right) [1-(-1)^{l-k}\cos \left( \triangle m\,\triangle
t\right) ],\triangle t=\left| t_{1}-t_{2}\right| .  \label{8}
\end{equation}
Hence we may get eq.$\left( \ref{1}\right) $ via the equality which defines
the conditional probability reported in the commented paper\cite{Go} in
terms of the joint probability, namely 
\[
R_{i}=\tau \exp (2t_{i}/\tau )r_{kl},\;j=\left| k-l\right| +1, 
\]
where $i=1(2)$ if particle $1(2)$ is the one decaying first. This proves
that our LHV model\'{}s prediction agrees with the quantum one for the said
experiment.

The model may be interpreted physically saying the either particle produced
in the decay of the Y(4S) oscillates between the two flavour states in such
a way that the flavours or the two particles in a pair are opposite at equal
proper times. The model looks somewhat contrived due to the lack of
symmetry, in the sense that the functions $P_{k}$ are quite different from
the functions $Q_{l}$. A more symmetrical model may be obtained assuming
that the assignement of the functions $P_{k}$ and $Q_{l}$ to the particles
in a pair is at random. In any case our purpose was only to show the
compatibility of the experiment with local realism, and not to make a
physically plausible model.

I acknowledge useful comments by Albert Bram\'{o}n and Alberto Ruiz.

\end{document}